\documentclass[12pt]{article}
\usepackage{times}
\title{Passive optical time-of-flight for Non line-of-sight localization} 

\author
{Jeremy Boger-Lombard $^{1}$ and Ori Katz$^{1\ast}$\\
\\
\normalsize{$^{1}$Applied Physics Department, The Hebrew University of Jerusalem, }
\normalsize{Jerusalem, 9190401, Israel}\\
\\
\normalsize{$^\ast$E-mail:  orik@mail.huji.ac.il}
}
\date{}

\usepackage{graphicx} 
\usepackage{gensymb}  
\usepackage{float}    
\usepackage{textcomp} 
\usepackage{verbatim} 
\usepackage{color} 

\begin{document} 

\baselineskip24pt
\maketitle 
\begin{abstract}
Optical imaging through diffusive, visually-opaque barriers, and around corners is an important challenge in many fields, ranging from defense to medical applications.
 Recently, novel techniques that combine time-of-flight (TOF) measurements with computational reconstruction, have allowed breakthrough imaging and tracking of objects hidden from view. 
These light detection and ranging (LiDAR)-based approaches, however, require active short-pulsed illumination and ultrafast time-resolved detection.
Here, bringing notions from passive RADAR and passive geophysical mapping approaches, we present an optical TOF technique that allows to passively localize light sources and reflective objects through diffusive barriers and around corners. 
Our approach retrieves TOF information from temporal cross-correlations of scattered light, providing temporal resolution that surpasses the state-of-the-art ultrafast detectors by three orders of magnitude. 
We demonstrate passive localization of multiple white-light sources and reflective objects hidden from view, using a simple setup, with interesting potential for covert imaging.  
\end{abstract}

\section*{Introduction}
In recent years, there have been great advancements in the development of techniques that enable non-line of sight (NLOS) optical imaging for a variety of applications, ranging from microscopic imaging through scattering tissue to around-the-corner imaging \cite{freund1990looking,wang1991ballistic,kirmani2009looking,buttafava2015non,gupta2012reconstruction,heide2014diffuse,velten2012recovering,satat2017object,gariepy2016detection,chan2017non,o2018confocal,katz2012looking,bertolotti2012non,katz2014non,takasaki2014phase,edrei2016optical,bouman2017turning}.
The enabling principle behind these techniques is the use of scattered light for computational reconstruction of objects that are hidden from direct view.  This has been achieved in a variety of approaches, such as
wavefront shaping\cite{katz2012looking}, inverse-problem solutions based on intensity only imaging\cite{bouman2017turning,klein2016tracking}, 
speckle correlations\cite{bertolotti2012non,katz2014non}, and time-resolved measurements \cite{kirmani2009looking,buttafava2015non,gupta2012reconstruction,heide2014diffuse,velten2012recovering,gariepy2016detection,chan2017non}. While wavefront-shaping approaches allow diffraction-limited resolution, they require prior access to the target position or long iterative optimization procedures. 
Inverse-problem solutions based on intensity-only imaging do not require prior mapping of the scattering barrier, but suffer from a drastically lower resolution, dictated by the dimensions of the diffusive halo. 

Considerably higher resolution was obtained using speckle correlations \cite{bertolotti2012non,katz2014non,takasaki2014phase,edrei2016optical} or time-of-flight (TOF) \cite{kirmani2009looking,buttafava2015non,gupta2012reconstruction,heide2014diffuse,velten2012recovering,satat2017object,gariepy2016detection,chan2017non,o2018confocal} based approaches. The former rely on 'memory-effect' angular correlations of scattered light \cite{freund1988memory}, and allow diffraction-limited, single-shot passive imaging, using a simple setup. However, memory-effect based approaches suffer from a very limited field of view (FOV), are limited to planar objects, and to a narrow spectral bandwidth. 
TOF based approaches have recently allowed three-dimensional (3D) tracking and reconstruction of macroscopic scenes hidden from view\cite{velten2012recovering,satat2017object,gariepy2016detection,o2018confocal}. 
These approaches utilize the principle of light detection and ranging (LiDAR) to obtain 3D spatial information from temporal measurements of reflected light. This is achieved by using ultrafast detectors to measure the time it takes a short light pulse to travel from an illumination point on the diffusive barrier, to the target object and back to the barrier.
The scene is then computationally reconstructed from multiple TOF measurements at different spatial positions on the barrier. 

Intuitively, one may assume that in order to measure \textit{time} of flight, a pulsed source is mandatory, as is done in most common LiDAR, SONAR and RADAR systems. However, short pulsed illumination is not a fundamental requirement: it is possible to obtain high-resolution temporal information from cross-correlation of ambient broadband noise, without any active or controlled source. This principle is put to use in helioseismology \cite{duvall1993time}, ultrasound \cite{weaver2001ultrasonics}, geophysics \cite{snieder2002coda}, passive RADAR \cite{davy2013green}, and recently in optical studies of complex media \cite{badon2015retrieving}. Here, we adapt these principles of passive correlations imaging for 3D localization of hidden broadband light sources and reflective objects through diffusive barriers and around a corner. 
We retrieve high resolution TOF information from scattered light using a simple, completely passive setup, based on a conventional camera.  
In our approach, temporal cross-correlations of scattered light are measured in a single shot, via low-coherence (white-light) interferometry, using controllable masks. 

Unlike conventional active TOF/LiDAR, where the temporal resolution is dictated by the pulse duration, or by the detectors' response time, in our approach the temporal resolution is dictated by the coherence time of the scene illumination.
For natural white-light illumination, as used in our experiments, the TOF temporal resolution is $\sim10 fs$,  three orders of magnitude better than the state of the art ultrafast detectors \cite{gariepy2016detection,velten2012recovering,o2018confocal} .

\section*{Principle} 
The principle of our approach is described in Fig. 1. Consider a small light source, or a reflective object hidden behind a diffusive barrier (Fig. 1a).
For a source transmitting short pulses (or an object reflecting a short pulse), the source position can be determined by measuring the time of arrival of light from the source to different points on the barrier. 
Such TOF-based spatial localization is straightforward, as was recently demonstrated using ultrafast detectors \cite{gariepy2016detection,velten2012recovering,o2018confocal}. 
However, when the illumination source is an uncontrolled continuous broadband noise source, such as natural white light sources, measuring the relative TOF is still surprisingly possible, by temporally cross-correlating the random time-varying fields arriving at the barrier (Fig. 1b,c). Such passive 'correlation-imaging' approach \cite{garnier2016passive} is the underlying principle of our approach. 

A numerical example for this principle is shown in Fig. 1a-c: the random time-varying fields from a broadband white-light source are measured at two positions on the barrier by two detectors (Fig. 1a). For a single hidden point source the measured fields arriving on the barrier, $E_1(t)$ and $E_2(t)$ are identical delayed versions of the source random field $E_s(t)$: $E_1(t)=E_s(t-\tau_1)$, and $E_2(t)=E_s(t-\tau_2)$  (Fig. 1b). Assuming free space propagation between the source and the barrier, 
 $\tau_{i} =  L_{i} / c$ , where $L_i$ is the optical path length between the source and the $i$-th measurement point, and $c$ is the speed of light. The TOF difference $\Delta \tau =  \Delta L / c = (L_1-L_2)/c$, can be determined by temporal cross-correlating the two arriving random fields (Fig. 1c). For a sufficiently thin scattering barrier, the cross-correlation of the measured fields exiting the barrier is identical to the cross-correlation of the arriving fields (See Supplementary section 8).   
The source position can be determined in three-dimensions from three or more such TOF measurements taken at different points on the barrier, in a similar manner to the principle of GPS, and in the recent NLOS imaging works in optics \cite{gariepy2016detection,velten2012recovering,o2018confocal}. 

The spatial localization accuracy is dictated by the TOF temporal resolution, $\delta t$,  i.e. by the temporal width of the cross-correlation peak. For a broadband source, this width is the source coherence time  $\delta t \approx t_c \approx 1/\Delta f$, where $\Delta f$ is the source spectral bandwidth. This is easily shown by noting that the cross-correlation of the two fields, $E_1 (t)\star E_2(t)$, is the autocorrelation of the source field, shifted by  $\Delta \tau $:  
\begin{equation}
E_1 (t)\star E_2(t)=(E_s\star E_s )(t-\Delta \tau)
\label{eq1}
\end{equation}
For a broadband source, the autocorrelation $(E_s\star E_s)(t)$, is a narrow, sharply peaked function around $t=0$, with a temporal width that is given by $\delta t \approx 1/\Delta f$, according to the Wiener-Khinchin theorem. 
Thus, the fields cross-correlation would display a sharp peak at the time delay $t=\Delta \tau=(L_1-L_2)/c$, providing the TOF difference from the source to the two points on the barrier. 
Such a single TOF difference measurement localizes the source to be on a hyperboloid surface. For a distant source, this provides the direction of arrival (DOA), as is exploited e.g. by the human brain for sound source localization. Repeating this passive TOF measurement at two additional positions would localize the source in 3D. 

Direct implementation of such a field correlation approach with two optical detectors is not straightforward, since measuring the temporal variations in optical fields requires phase-sensitive ultrafast detection with sub-femtosecond temporal response. Conventional optical detectors measure only the optical intensity, averaged over response times that are several orders of magnitude longer. However, access to the optical fields' temporal cross-correlation is directly possible via interference, even when using slow detectors. This is the principle of our optical implementation for passive TOF, presented in Fig. 1d-g, and explained below. 

Fig. 1d depicts the optical setup that enables spatially-resolved temporal cross-correlation measurements via interferometry, using a conventional camera. 
First, the light from two chosen positions on the barrier is selected by a mask having two apertures. The mask is placed adjacent to the barrier or imaged on its surface. The light fields from the two apertures, $E_1(t)$ and $E_2(t)$, interfere on a camera placed at a sufficiently large distance from the mask. 
In a single, long exposure image, each camera pixel value, $I(x)$, at a transverse position $x$, is the time-integrated optical intensity resulting from the interference of the two fields:  
\begin{equation}
I(x)=\int _{-\infty}^{\infty}|E_1 (t)+E_2 (t+t(x) )|^2 dt=I_1+I_2+2[E_1\star E_2](t(x))
\label{eq2}
\end{equation}Thus, a single camera row provides the fields temporal cross-correlation, $E_1\star E_2 $,  sampled at thousands  of different time delays, given by the positions of the camera pixels, $x$: $t(x)=\Delta L(x)$. The sampled time-delays are dictated by the system's geometry (see Supplementary fig. S6).

The source TOF difference can be easily determined from the position of the cross-correlation peak, manifested as low-coherence interference fringes (Fig. 1f,g). 
This bears similarity to  white light interferometry \cite{hausler1996observation} and also optical coherence tomography (OCT) \cite{huang1991optical}. However, here, no reference arm or external source are used, and the measurements are self-referenced.

In practice, for a strongly scattering barrier, the light intensity from each of the apertures on the mask produces a random speckle intensity pattern on the camera (Fig. 1e). 
Importantly, the speckle intensity patterns do not mask the low-coherence interference fringes, as we design the fringes period to be considerably smaller than the speckle grain size  (Fig. 1f,g, see also Supplementary fig. S3).
\renewcommand{\figurename}{Fig.}
\begin{figure}[H]
\centering
\includegraphics[width=12cm]{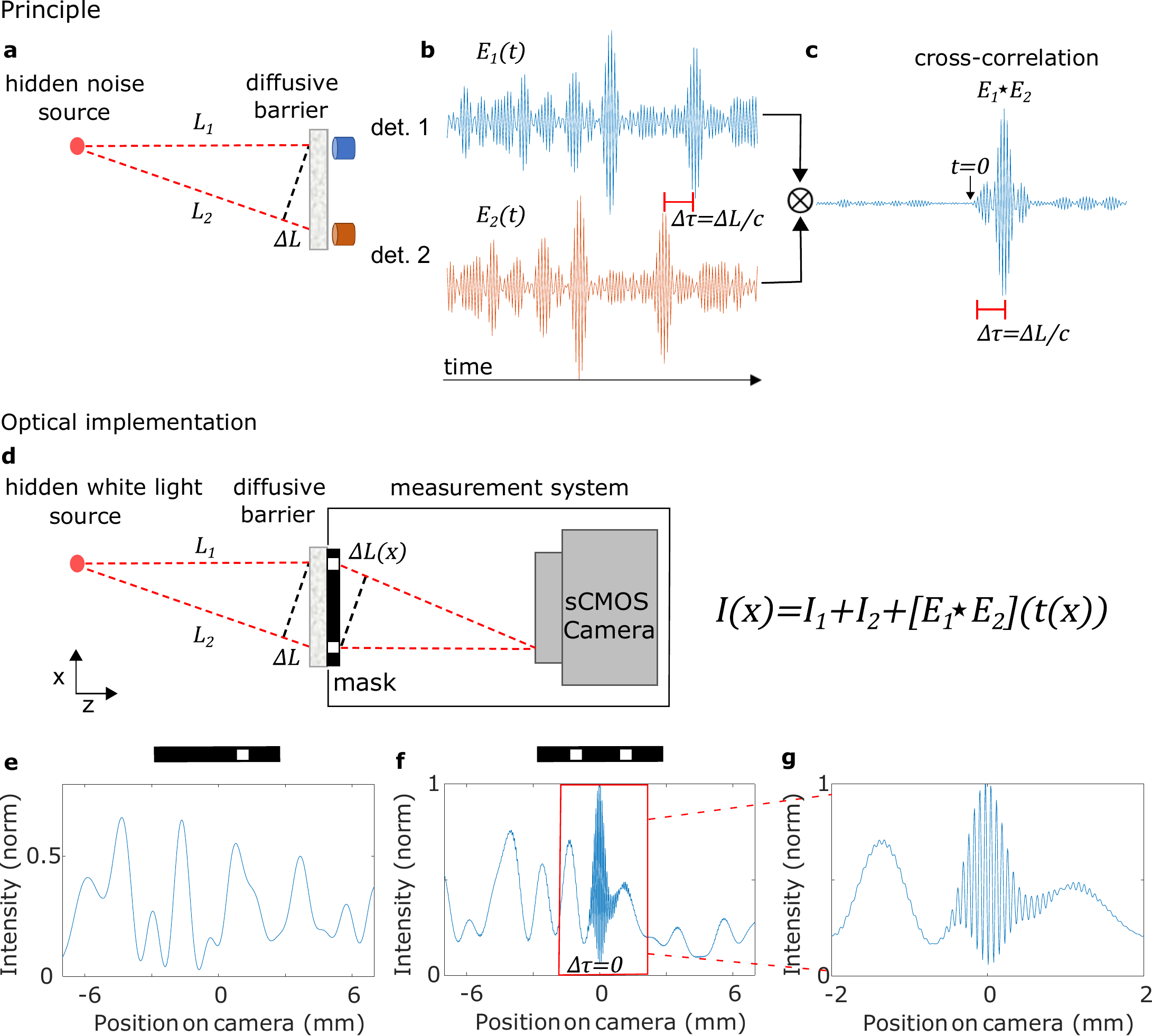}
\caption{\textbf{Passive TOF through diffusive barriers by temporal cross-correlations. }(\textbf{a-c}) Principle: a white-light source or reflective object is hidden behind a diffusive barrier. The source random noise fields are measured at two points on the barrier by detectors 1 and 2. \textbf{(b)} For a thin barrier, the measured fields are replicas of the same broadband noise, shifted by the TOF difference $\Delta \tau = \Delta L / c$.  \textbf{(c)} Cross-correlating the measured fields reveals the TOF difference as a sharp peak. 
\textbf{(d)} Optical implementation: light from two points on the barrier is selected by a controllable mask, and interferes on a high resolution camera. A single long-exposure camera shot $I_{cam}(x)$ provides the fields temporal cross-correlation at different time delays. \textbf{(e-g)} Numerical example of the measured intensity patterns on a single camera-row: \textbf{(e)} For a mask with a single aperture, a speckle intensity pattern is measured. \textbf{(f)} For a double-aperture mask, a cross-correlation peak appears as white-light interference fringes on top of the speckle pattern, providing the TOF. \textbf{(g)} Zoom-in on (f), the TOF resolution is the source coherence-time.} 
\label{fig:fig1}
\end{figure}
\section*{Results}
\subsection*{3D passive TOF measurement through a diffusive barrier}
Fig. 2 presents experimental results of passive TOF measurement using a setup based on the design of Fig. 1d. The full experimental setup is given in Supplementary Figure S1.  As a first demonstration, a single small white light LED source was hidden behind a highly scattering diffusive barrier, with  $80\degree$ scattering angle, having no ballistic component ($80\degree$ light shaping diffuser, Newport). A movable mask, comprised of four small apertures (Fig. 2d), is used to interfere light from two pairs of points on the barrier. The points are vertically and horizontally spaced to obtain TOF information on both elevation and transverse position of the hidden source, respectively (see below). 

When a mask with only a single aperture is used, the camera image is a random speckle pattern (Fig. 2a).
This random speckle pattern provides no useful spatial information on the source position. However, when a mask with two horizontally-spaced apertures is used, low-coherence interference fringes appear on top of the random speckle pattern at a specific horizontal position on the camera image (Fig. 2, b and c). This white-light fringe pattern is a result of the interference of the two speckle patterns that are transmitted through each of the two apertures. Since the hidden source is a broadband white-light source, the interference fringes are located only around the zero path delay difference, i.e. when the path difference accumulated after the diffuser $ \Delta L(x)$ is equal to the path difference accumulated before the diffuser:  $ \Delta L(x)=\Delta L$  (Fig. 1d). Thus, the optical path length (or TOF) difference of the light from the source to the two apertures can be extracted from the fringes position in a single camera image (Fig. 2c), with a resolution given by the coherence-length (or coherence-time) of the source, and the additional path delay spread inside the diffusive barrier (see Discussion, and Supplementary section 8).  
\begin{figure}[H]
\centering
\includegraphics[width=10cm]{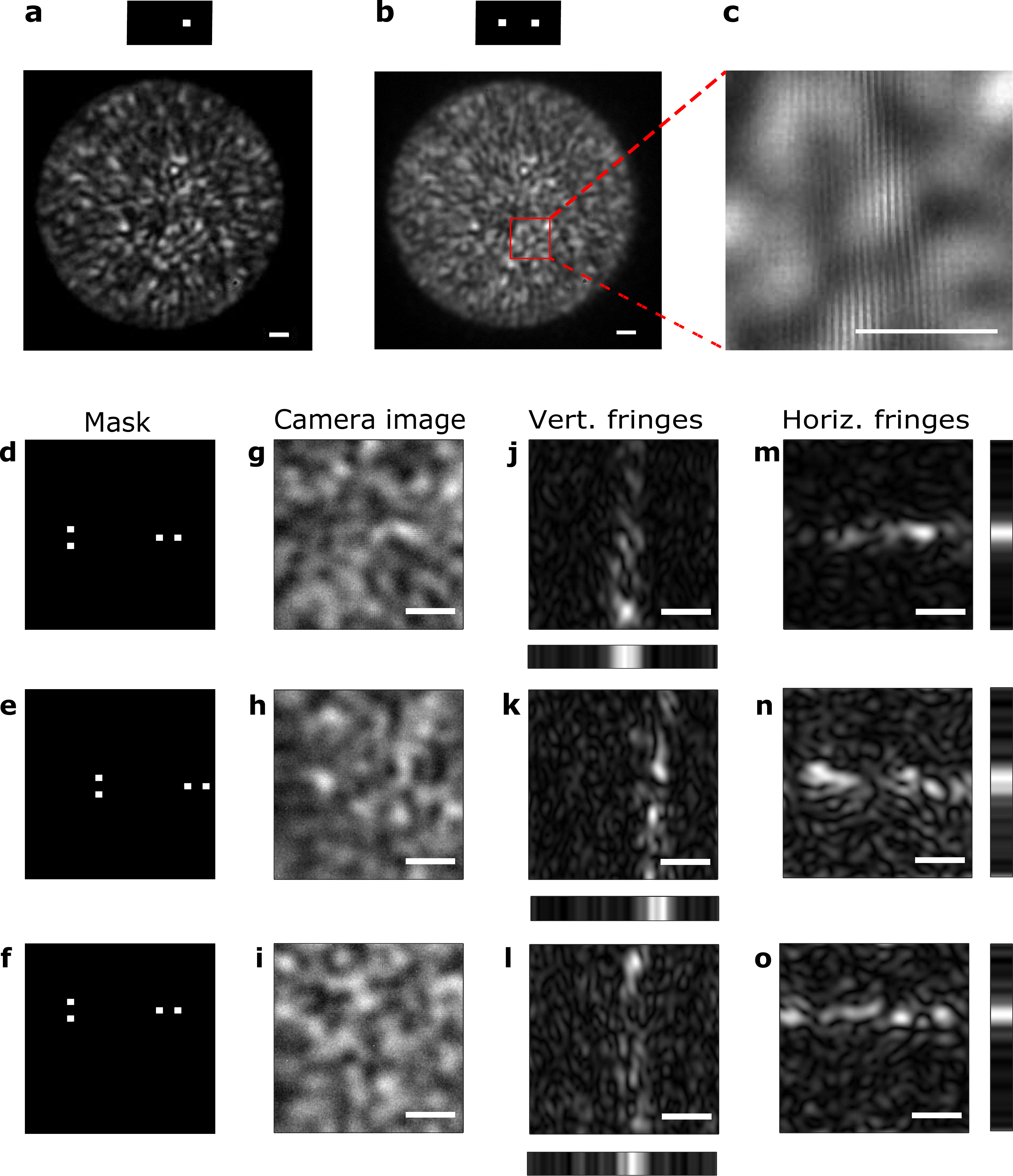}
\caption{\textbf{Experimental passive TOF measurement of a single hidden source through a diffusive barrier.}\textbf{ (a-c)} Raw camera images showing: \textbf{(a)} The speckle pattern measured for a mask with a single aperture. \textbf{(b)} For a double-aperture mask (top), localized interference fringes (red square) provide the TOF difference from the hidden source to the two apertures. \textbf{(c)} Zoom-in on (b).  \textbf{(d-f)} Shifting the mask for multiple TOF measurements, allow localization of the source. For 3D localization, elevation information is obtained by multiplexing a vertically spaced double aperture. 
\textbf{(d-o)} Example for TOF retrieval: \textbf{(d-f)} Mask shifted horizontally (e) and vertically (f) as imaged on the barrier. (\textbf{g-i}) Raw camera images for each mask position. (\textbf{j-l}) Vertical fringes envelope as detected by filtered Hilbert transform
(bottom: sum over rows). (\textbf{m-o}) Horizontal fringes envelope, providing the TOF difference between the two vertically-separated apertures (right: sum over columns). 
Scale bars, 92fs.}
\label{fig:fig2}
\end{figure}

A single optical path difference measured from a single camera image, localizes the source position to be on a hyperboloid surface, with its foci at the two apertures. To localize the source to a single point,  additional measurements are required. Two additional measurements with the mask placed at two different positions will provide two additional hyperboloids. The intersection of the three hyperboloids can localize a single source to a single point in 3D.

Inspired by aperture masking interferometry \cite{haniff1987first,tuthill2000michelson}, we designed a mask that multiplex two TOF measurements in a single camera shot. This mask, presented in Fig. 2d, comprises two pairs of apertures, simultaneously providing two TOF measurements by generating both vertical and horizontal high spatial-frequency fringes. Fig. 2g presents an example for a raw camera image acquired using this mask. 

The position of the high spatial-frequency horizontal and vertical fringes can be easily determined using spatial bandpass filtering and a Hilbert transform, or spectrogram analysis (see Supplementary fig. S3). The results of such filtering on the camera image of Fig. 2g are shown in Fig. 2j,m, revealing the vertical and horizontal fringes respectively. Additional TOF measurements are obtained by shifting the mask position, horizontally (Fig. 2e,h), and vertically (Fig. 2f,i). When the mask is horizontally shifted only the horizontal fringes position changes (Fig. 2k), due to the change in TOF, while the vertical fringes position remains fixed (Fig. 2n). When the mask is vertically shifted, the horizontal fringes shift upwards (Fig. 2o),  and the vertical ones remain fixed (Fig. 2l).

\subsection*{Localization of multiple light sources}
In the case of multiple hidden sources, each single camera image contains several interference fringe patterns at different positions. Each fringe pattern position provides the TOF for a single hidden source. For independent spatially incoherent sources, such as natural light sources, the number of interference fringe patterns is equal to the number of hidden sources, as the light from different sources does not interfere. 

To localize multiple hidden sources without ambiguity, a larger number of TOF measurements is required. This can be achieved by shifting a single mask to multiple positions, and identifying the fringes for each of the mask's positions. An experimental example for such localization of two and three hidden sources in two dimensions is shown in Fig. 3. Fig. 3b displays the fringes envelope measured at 40 different mask positions: Each row in Fig. 3b displays the fringe amplitude extracted from a single camera image (see Supplementary fig. S3), where the horizontal axis is the TOF delay (camera pixel), and the vertical axis is the mask position. Inspection of Fig. 3b clearly reveals that two hidden sources are present at the hidden scene.



Each intensity-peak in Fig.3b localizes the sources to a hyperboloid. The sources are unambiguously localized by intersecting the multiple measured hyperboloids, as is demonstrated in Figure 3c. An example of experimental reconstruction a scene with three hidden sources is presented in Figure 3d. The sources positions can also be alternatively determined from the slope and positions of each high brightness line in Fig.3b\cite{liu20153d}.

In the experiments shown in Fig. 3, the mask position was mechanically scanned. A similar acquisition can be performed without any moving parts, by replacing the mask with a programmable dynamic mask, implemented using a computer-controlled spatial light modulator (SLM). An experimental demonstration using such a programmable mask for passive TOF measurements is presented in Supplementary fig. S2. The advantages of an SLM-based mask are its versatility and speed, in particular for advanced multiplexing (see Supplementary section 2).   

\begin{figure}[H]
\centering
\includegraphics[width=12cm]{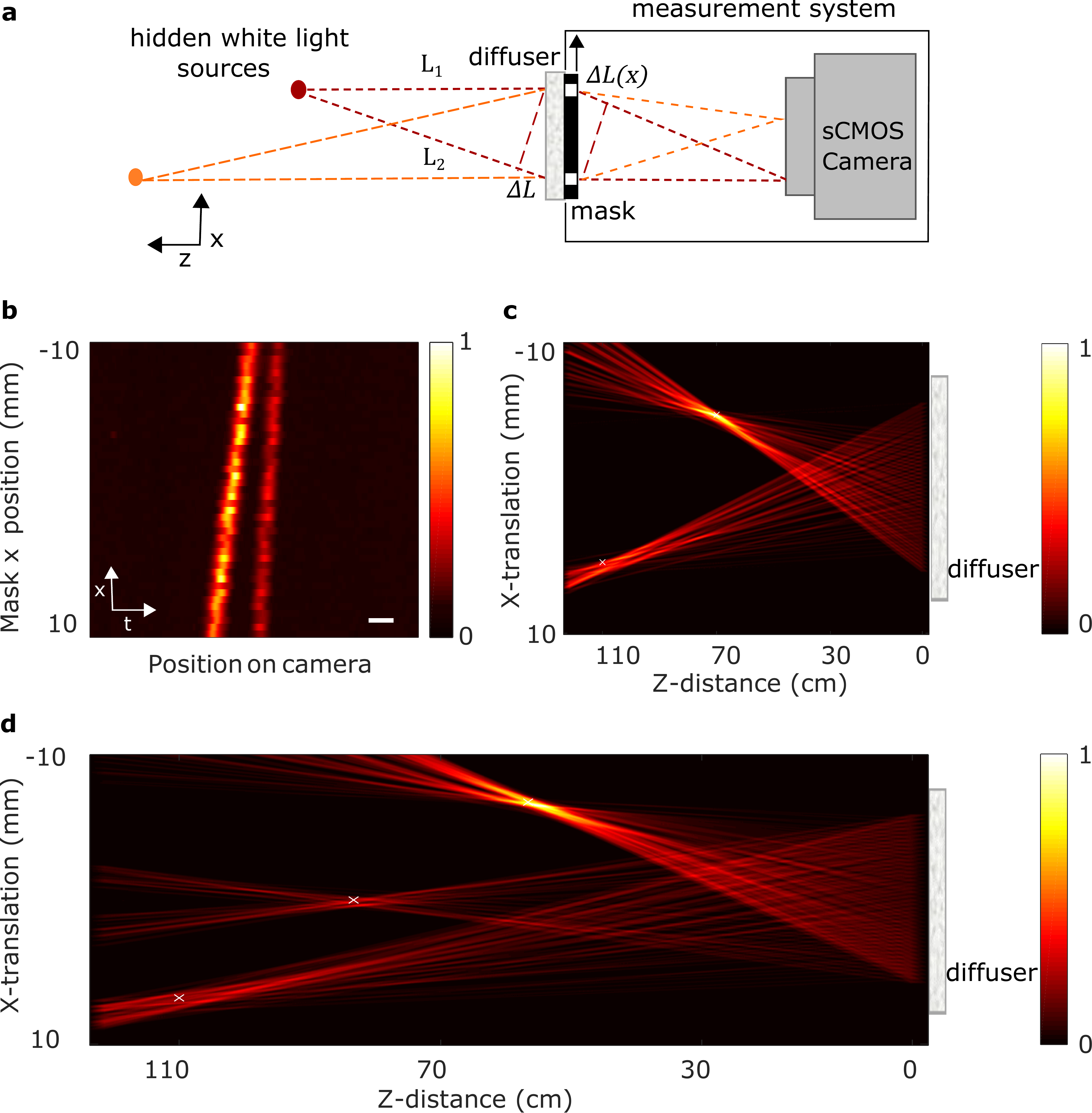}
\caption{\textbf{Experimental localization of multiple hidden sources:} \textbf{(a)} Setup with two hidden sources (simplified depiction).  \textbf{(b)} Detected envelope of the interference fringes as a function of the mask position (vertical axis). The fringes positions (horizontal axis) mark the TOF differences.
\textbf{(c)} Hidden scene reconstructed from (b), where each fringe position in (b) localizes the sources to a hyperboloid. The sources real positions are marked by 'x' .\textbf{(d)} Example of reconstruction of a scene containing three hidden sources. Scale bar, 92fs.}
\label{fig:fig3}
\end{figure}

\subsection*{Localization around a corner}
Our passive TOF approach can be used to localize sources 'around a corner', using light that was scattered off diffusive white-painted walls \cite{velten2012recovering,gariepy2016detection,o2018confocal}. Fig. 4 presents a proof of principle experiment for around the corner passive localization. Fig. 4a depicts the experimental arrangement, which is conceptually identical to the design of Fig. 2a, with the difference that the diffusive barrier is replaced by a white painted wall, and that the wall's surface is imaged on the movable mask.  

In order to localize multiple light sources that are placed 'around the corner', the same measurement protocol described above (Fig. 3) was performed. The resulting spatio-temporal (x-t) trace of fringes' position on camera (i.e. TOF, t) vs. mask position (x)  is shown in Fig. 4c. The reconstructed scene is shown in Fig. 4d.

Fig. 4e shows a comparison of the fringes envelope as a function of the time delay, for the case of a thin diffuser, and the white painted barrier. As a result of multiple-scattering in the white-paint, the reflective barrier behaves as an effective thick scattering medium, having a temporally broader impulse response. Thus, considerably broadening the fringes envelope, i.e. lowering the TOF resolution, and limiting the localization accuracy. Nonetheless, our approach is still able to successfully localize the sources positions.

\begin{figure}[H]
\centering
\includegraphics[width=12cm]{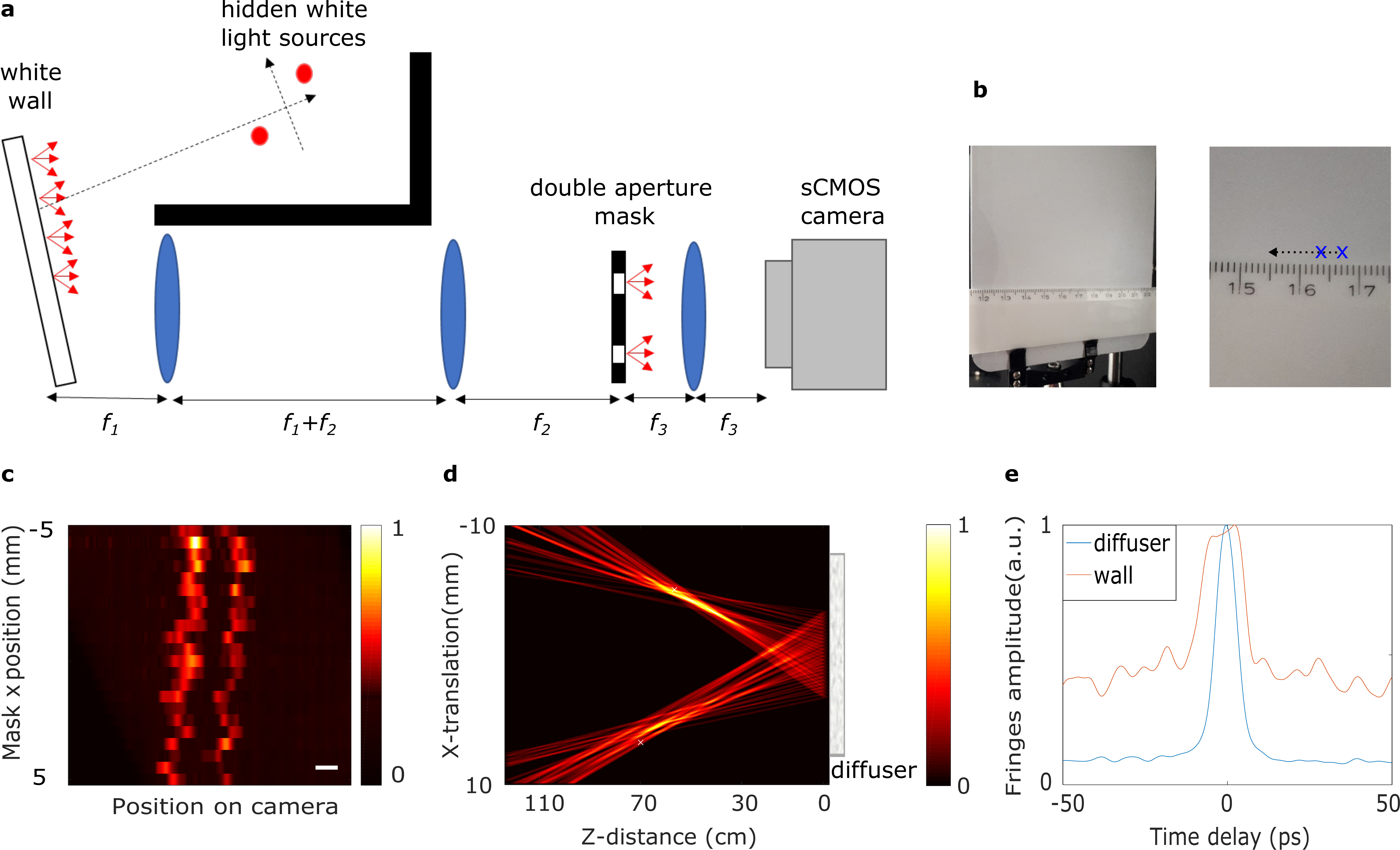}
\caption{\textbf{Localizing multiple light sources around a corner.} \textbf{(a)} Setup (top view): hidden white light sources illuminate a white-painted wall. The wall surface is imaged by a 4-f telescope on a movable double-aperture mask. The light from the two apertures is interfered on the camera by a lens. \textbf{(b)} Images of the wall surface. The apertures positions are marked by 'x'. The scan trajectory is depicted by a dashed line. \textbf{(c)} Interference fringes envelope as a function of the double-aperture mask position, revealing the two hidden sources. \textbf{(d)} Scene reconstructed from (c). The sources real positions are marked by 'x'. \textbf{(e)} fringes envelope as a function of the time delay, for the case of a thin diffuser, and the "thick" white painted barrier.  Scale bar: 92fs.}
\label{fig:fig4}
\end{figure}

\subsection*{Localization of reflective objects}
The approach can be used to localize reflective objects. Fig. 5 presents such a demonstration using  the same experimental system, with the only difference that two small metallic objects are placed in the scene, and the scene is illuminated by a Halogen lamp (Thorlabs OSL2). The measured spatio-temporal (x-t) TOF trace and scene reconstruction are presented in Fig. 5b-c.

\begin{figure}[H]
\centering
\includegraphics[width=12cm]{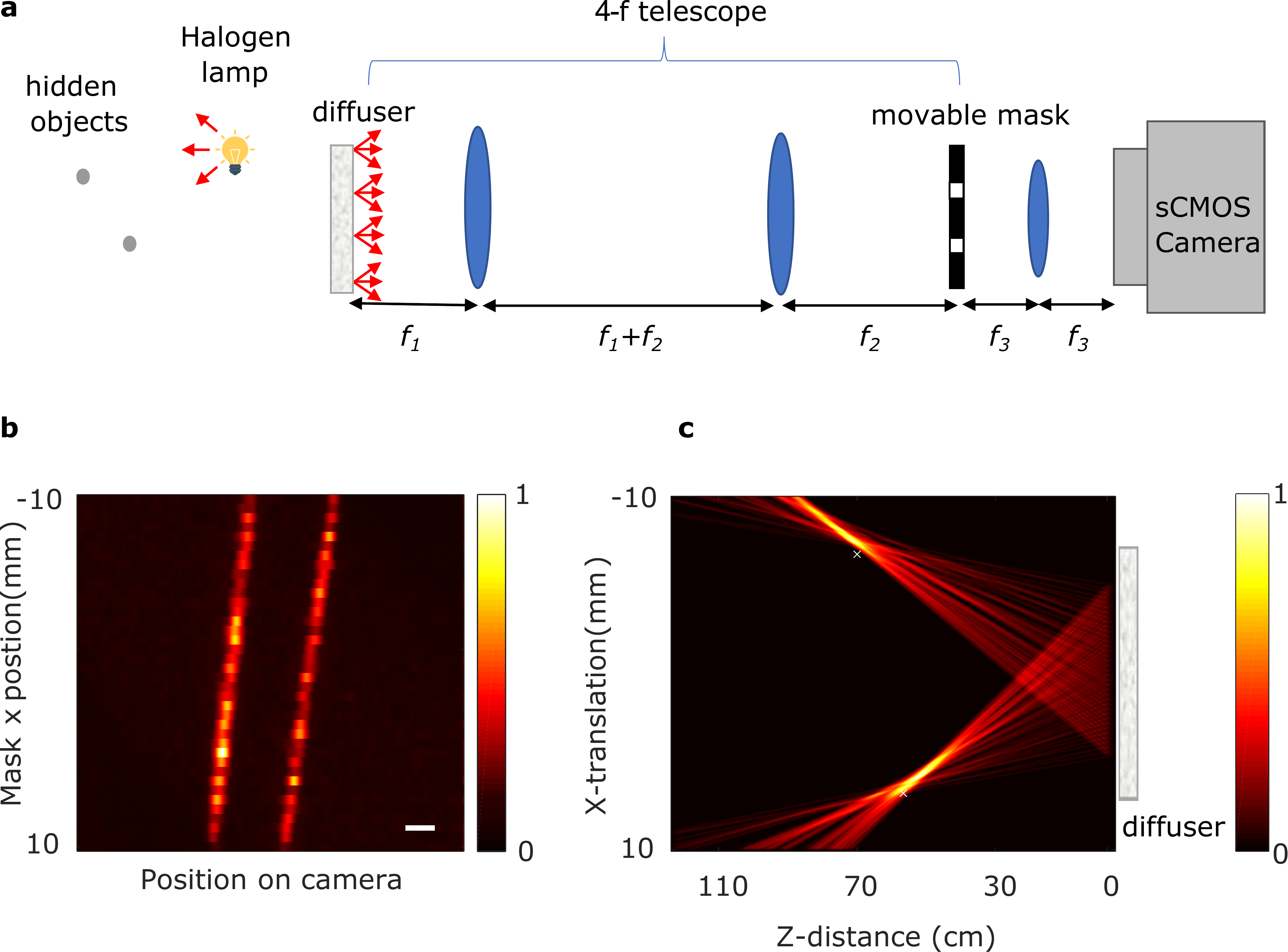}
\caption{\textbf{Localizing hidden reflective objects.} \textbf{(a)} Setup (top view): Two hidden metallic objects are illuminated by a Halogen lamp (Thorlabs OSL2). The reflected light passing through a highly-scattering diffuser is interfered on a camera by a movable double-aperture mask. \textbf{(b)} Interference fringes envelope as a function of the mask position, providing the TOF information. \textbf{(c)} Scene reconstructed from  (b) revealing the sources positions. The real positions are marked by 'x'. Scale bar: 92fs.}
\label{fig:fig5}
\end{figure}

\section*{Discussion}

We have demonstrated an approach that allows to passively localize incoherent light sources and reflective objects, through diffusive barriers and around corners.
As in conventional (active) TOF approaches \cite{velten2012recovering,gariepy2016detection,o2018confocal}, the spatial localization resolution is determined by the TOF temporal resolution and the setup's geometry. However, unlike conventional TOF approaches, where the temporal resolution is dictated by the detectors response time, in our approach the TOF resolution is given by the light source coherence-time. For the white light sources used in our experiments, the coherence time is of the order of $10fs$, more than three orders of magnitude better than the response time of state-of-the-art streak cameras or SPAD detectors \cite{velten2012recovering,gariepy2016detection,o2018confocal}. This temporal resolution may be an advantage in microscopic imaging applications.

Beyond the TOF resolution, the main advantage of our approach is in it being passive and having a simple compact implementation, not requiring fast detectors or streak cameras. Our technique makes use of natural light, present in many natural scenes in a fundamentally new fashion. 

However, the passive nature of our TOF approach is also its main disadvantage: since light from natural light sources is used, the detected intensity levels are inherently low. In our implementation, integration times of a few seconds were required in the diffusive barrier case. For around the corner localization exposure times of ~15 minutes were required (Fig. 4). More advanced signal acquisition, e.g. multiplexing \cite{tuthill2000michelson}, and reconstruction algorithms should allow reduced acquisition times.

The presented technique is conceptually similar to Michelson stellar interferometry, where the fringe contrast (visibility) as a function of the apertures separation is used to reconstruct the shape of a distant star., as was recently demonstrated for around-the-corner object reconstruction \cite{batarseh2018passive}. However, in our approach the apertures separation is fixed, and only the interference fringes positions are used to extract the TOF information. Our approach can be combined with such stellar-interferometry approaches to reconstruct the hidden source's shape. Interestingly, the same information that was obtained here via interferometry can be obtained from intensity only correlations, using Hanbury Brown and Twiss (HBT) interferometry \cite{brown1956correlation}.

To obtain high-contrast fringes the aperture separation must be smaller than the source coherence size on the barrier, $r_{coh}$. This limits the apertures separation when large extended sources or objects are considered. 
Since larger aperture separation (base-length) is desirable for better spatial localization resolution, 
the localization resolution is worse for larger objects. An analysis performed in Supplementary Section 6  shows that the angular resolution from a single camera image is of the order of the object angular extent. The final localization accuracy is considerably improved as it is obtained from multiple TOF measurements at different masks positions.

In our experiments with thin diffusive barriers and highly scattering walls the fringe visibility was high. However, when thick diffusive barriers are considered 
the fringe visibility, as well as speckle contrast, will be reduced due to the narrower speckle spectral correlation width, resulting from the larger spread in optical path\cite{curry2011direct,small2012spatiotemporal}.

We have demonstrated our approach on scenes containing only several small hidden sources or objects. For bright background scenes the major challenge for applying our method will be the fringe visibility, which are expected to be considerably lower due to the background bright signal. 


In our approach TOF differences from the hidden scene are measured, rather than the total round-trip TOF to the hidden scene and back, measured in active TOF approaches \cite{velten2012recovering,gariepy2016detection,o2018confocal} and LiDAR. This leads to localization of the hidden sources on spatial \textit{hyperboloids} in 3D, rather than ellipsoids \cite{velten2012recovering,gariepy2016detection}, or spheres \cite{o2018confocal} of previous works.



Unlike memory-effect based works\cite{katz2014non}, the FOV of our technique is not limited by the memory-effect, since each source position is measured independently. 




\section*{Materials and Methods}
The full experimental setups are presented in detail in Supplementary fig. S1. 
The hidden light sources were generated by splitting the light from a white light LED (Thorlabs MWWHF1) to four using a fiber bundle (Thorlabs BF42HS01), effectively producing white light sources of $200 \mu m$-diameter and numerical aperture of $NA=0.39$, having an average power of $\sim 1.5mW$ .  
The diffusive barrier  was a Newport light shaping diffuser with a scattering angle of 80 degrees. The scattering wall was a metal plate painted with white matte spray (Tambour 465-024). 
The light sources were placed at various positions with distances of 30-110cm from the diffuser and 56-70cm from the wall.
For  the object localization experiment the light source used was a white light halogen lamp (Thorlabs OSL2) culminated with its culmination package (Thorlabs OSL2COL). The object was a metallic nut covered with black tape leaving a reflective area 3mm high and 0.7mm wide. 
The diffusive barrier / wall was imaged on the aperture masks using a 4-f telescope. The light collection aperture diameter on the first lens was $5cm$. For the measurements shown in Fig. 2 and Fig. 5 and for around the corner measurements, the aperture was limited to $2.5cm$. 
The apertures masks were fabricated by drilling $0.25mm$ diameter holes in \textbf{$\sim 2mm $}-thick black Delrin plates. The separation between the apertures was $3.2mm$. 
The light passing through the mask was focused on an sCMOS camera (Andor Neo 5.5) with an $f=10cm$ lens in an f-f configuration.

\baselineskip=24pt

\section*{Acknowledgements }

This material is based upon work supported by the Defense Advanced Research Projects Agency under Contract No. HR0011-16-C-0027. O.K. acknowledges support from the Azrieli Faculty Fellowship, Azrieli Foundation.

\section*{Author contributions}

O.K. conceived the idea,  J.B.L. and O.K. designed the experiments and performed numerical simulations. J.B.L. conducted the experiments and analyzed the data. J.B.L. and O.K. wrote the manuscript. 

\section*{Competing financial interests} 
The authors declare no competing financial interests. 

\section*{Corresponding author}
Correspondance to Ori Katz orik@mail.huji.ac.il

\newpage
\renewcommand{\thefigure}{S\arabic{figure}}
\renewcommand{\figurename}{Fig.}
\setcounter{figure}{0} 
\setcounter{equation}{0}
\setcounter{page}{1}

\begin{center}
\begin{Large}
Non line-of-sight localization by passive optical time-of-flight\\
Supplementary Materials\\
\end{Large}
\end{center}
\begin{center}

\begin{large}
Jeremy Boger-Lombard$^1$ and Ori Katz$^{1*}$\\
\end{large}
\medskip
$^1$Applied Physics Department, The Hebrew University of Jerusalem,\\ Jerusalem, 9190401, Israel\\
\medskip
$^*$orik@mail.huji.ac.il
\end{center}

\section{Optical setup}
The full experimental setup for imaging through a diffusive barrier is given in Supplementary fig. S1. The light source used for the experiments was a white-light LED source (MWWHF1, Thorlabs), which was split into four by coupling to  a fiber bundle (BF42HS01 Thorlabs). The sources in all the experiments were 2-3 tips of the four fiber bundle ends. The sources were positioned behind a highly scattering diffuser (light shaping diffuser, $80\degree $ scattering angle, Newport).

On the other side of the diffusive barrier, a 4-f telescope ($f_1=150mm, f_2=180mm$) imaged the barrier's surface on a movable mask, comprised of two small apertures (for 2D localization), or two pairs of apertures, for 3D localization.  The mask was placed at the front focal plane of a lens ($f_3=100mm$). A cooled sCMOS camera (Andor Neo 5.5) placed at the back focal plane of the lens, records the diffraction pattern of the scattered light that is transmitted through the double aperture mask. 
For the measurements of localizing an object, the diffuser used was Newport light shaping diffuser with a scattering angle of $40\degree$.

\begin{figure}[H]
\centering
\includegraphics[width=12cm]{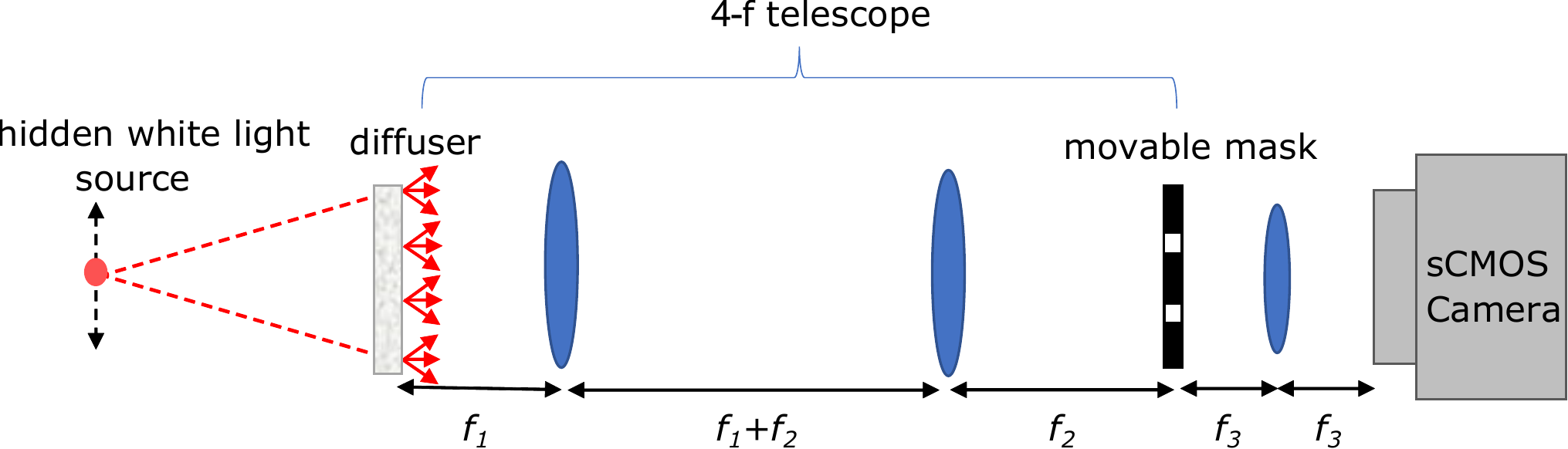}
\caption{\textbf{Setup for passive TOF through a highly scattering medium:} a white-light LED “point” source is hidden behind a diffusive barrier. A double aperture mask that is 4-f imaged on the barrier,  selects diffused light from two points on the  barrier to be interfered on an sCMOS camera. The mask is translated orthogonal to the optical axis to measure different points on the barrier.} 
\label{fig:sup1}
\end{figure}
\section{Results using a dynamic programmable mask }
In order to go beyond the limitations of a static double-aperture mask that is mechanically scanned across the barrier, the setup shown in Supplementary fig. S2a was constructed. In this system, the double aperture mask is replaced by a programmable "digital" amplitude mask constructed using a spatial light modulator (SLM). For amplitude shaping, the phase-only SLM (Holoeye Pluto BB) was placed between two linear polarizers in a cross-polarization configuration. 

The setup for a dynamic programmable mask is conceptually identical to the setup using a mechanical-scanned fix mask (Supplementary fig. S1), with two main technical differences:  The first is that in order to obtain a high contrast amplitude mask using the specific liquid crystal SLM model (Holoeye PLUTO BB), which had significant chromatic dispersion, a narrow bandpass filter (BPF) with a 10nm bandwidth (FB550-10 Thoralbs) was placed before the camera. This reduces the light utilization efficiency of this setup, and could be improved by using less dispersive SLMs, amplitude only SLMS, or potentially MEMS based SLMs.
The second difference of the programmable mask setup of Supplementary fig. S2a, was that the reflective SLM  was placed at a close distance ($<5cm$) to the diffusive barrier, instead of being 4-f imaged on it. Even with such imperfect conjugation, two light sources could be simultaneously separated and localized with our approach through a diffusive barrier, by displaying a double aperture mask at different positions on the SLM (Supplementary fig. S2b,c). In this experiment the diffusive barrier was a light shaping diffuser, with a scattering angle of 5 degrees (Newport).

Supplementary fig. S2b presents the results of fringe localization from 18 different positions of the displayed double mask aperture, achieved without any mechanical scanning: each row in Supplementary fig. S2b, is the fringe envelope amplitude at different positions of the camera, extracted using spectrogram analysis (see Supplementary fig. S3). The reconstructed sources positions from these measurements are shown in Supplementary fig. S2c. 

A programmable mask is advantageous over a mechanically scanned fixed mask as it does not require mechanical scanning, it can straightforward implement advanced multiplexing approaches that can be used for reconstructing the source image , as done in aperture masking interferometry\cite{tuthill2000michelson}. 

The main disadvantage of a programmable mask is the lower contrast and transmission compared to a mechanical mask, and potential chromatic aberrations. In our implementation the chromatic aberrations required the use of a narrow bandpass  filter, which not only resulted in lower light utilization efficiency, but also reduced the temporal resolution by an order of magnitude, to be of the order of $100 fs$, still considerably better than the response time of the state of the art detectors. The lower temporal resolution can be observed in a larger width of the curves shown in Supplementary fig. S2b. 

\begin{figure}[H]
\centering
\includegraphics[width=12cm]{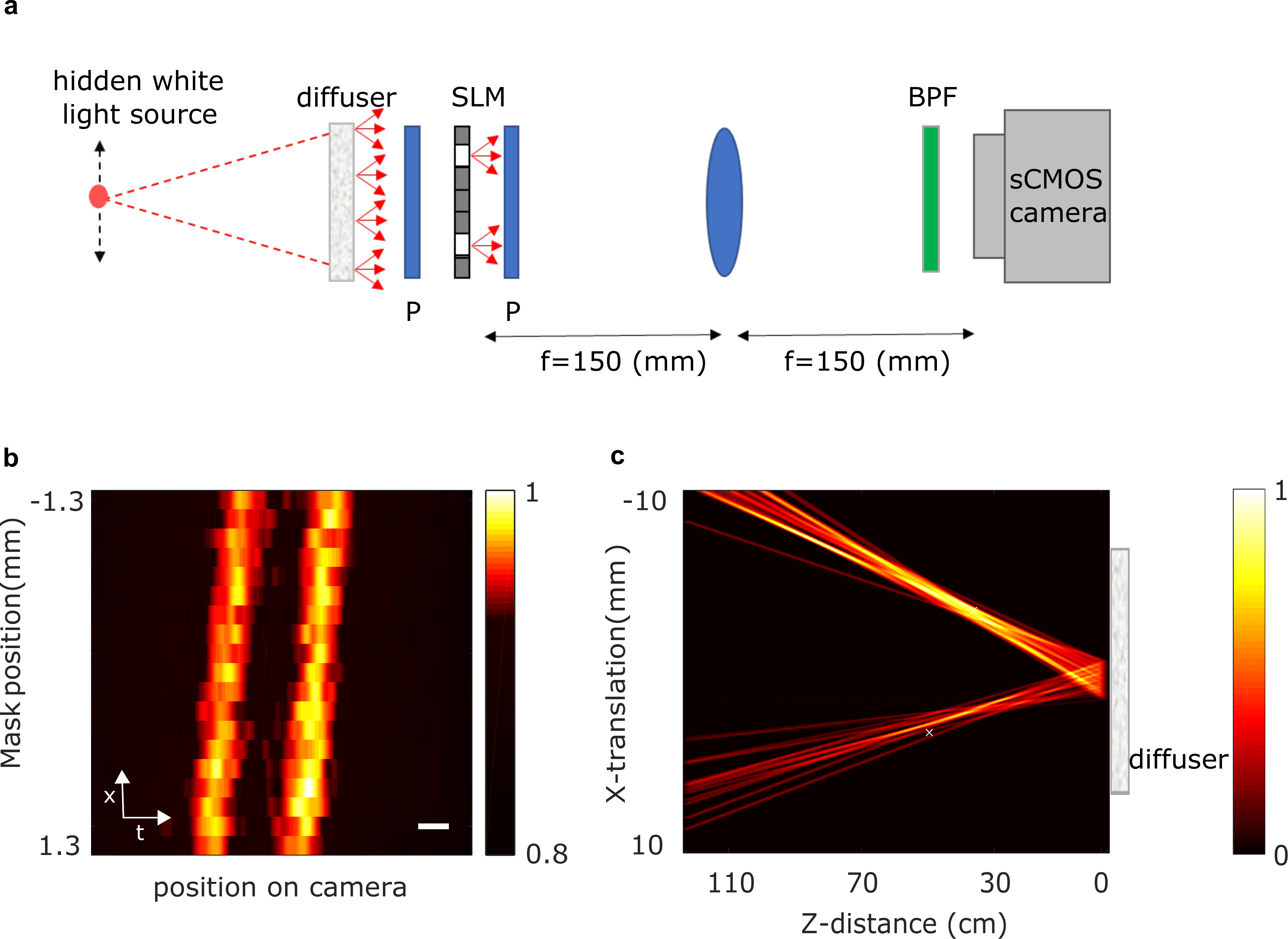}
\caption{|\textbf{Localization of multiple hidden sources using a dynamic programmable mask.} \textbf{(a)} Setup: a white light LED source is hidden behind a diffusive barrier, the light is collected from two points on the scattering barrier by displaying a double aperture amplitude mask on an SLM. The interference pattern is recorded on a camera (Zyla 4.2 plus). \textbf{(b)} Interference fringes envelope as a function of the mask position, providing the TOF information.\textbf{c} Scene reconstructed from (b) revealing the sources positions. The real positions are marked by 'x'. Scale bar: 92fs.}
\label{fig:sup2}
\end{figure}

\section{Constructing the spatio-temporal TOF (x,t) maps}
Supplementary fig. S3a shows a raw camera image taken with the setup of Supplementary fig. S1 . The image seems to be a random, low contrast, speckle pattern, as would be seen through a highly scattering barrier. However, as a result of the double aperture mask, low coherence (white-light) fringes are present in the image, at a position which reflects the TOF difference (Supplementary fig. S3a, red arrow). The fringes are localized around the zero delay time providing the optical path difference of the light from the hidden light source to the two apertures. 

In order to allow detection of the fringes, we made sure that the fringes period is considerably smaller than the speckle grain size. This was ensured by using double aperture masks having an aperture separation distance that is larger than the size of each aperture. Thus, the fringes can be localized by high-pass (or bandpass) filtering the camera image, around the fringes spatial frequency. One possible approach to perform such filtering is via a 2D Hilbert transform, as shown in Fig. 2.

Another equivalent processing approach is to perform a spectrogram analysis  for each camera row. The result of such an analysis is shown in Supplementary fig. S3b: The shown result is the average of spectrogram analysis performed over each of the camera image rows, after averaging each 10 camera rows. The spectrogram analysis performs a  short time Fourier transform (STFT) around each horizontal pixel position with a chosen window length (here, 128 pixels). This provides spatial frequency information to be analyzed with a spatial resolution of the window length. The direct result of such  simple spectrogram analysis shows a clear spectral peak that is localized around the position of the interference fringes. Taking the relevant spectrogram row (i.e spatial frequency of the interference fringes, Supplementary fig. S3b, dashed box)  provides the position of the fringes for this particular mask position. Repeating this process  for different mask positions provides the spatio-temporal TOF trace of Supplementary fig. S3c, which allows the source localization (here a single source).

\begin{figure}[H]
\centering
\includegraphics[width=16cm]{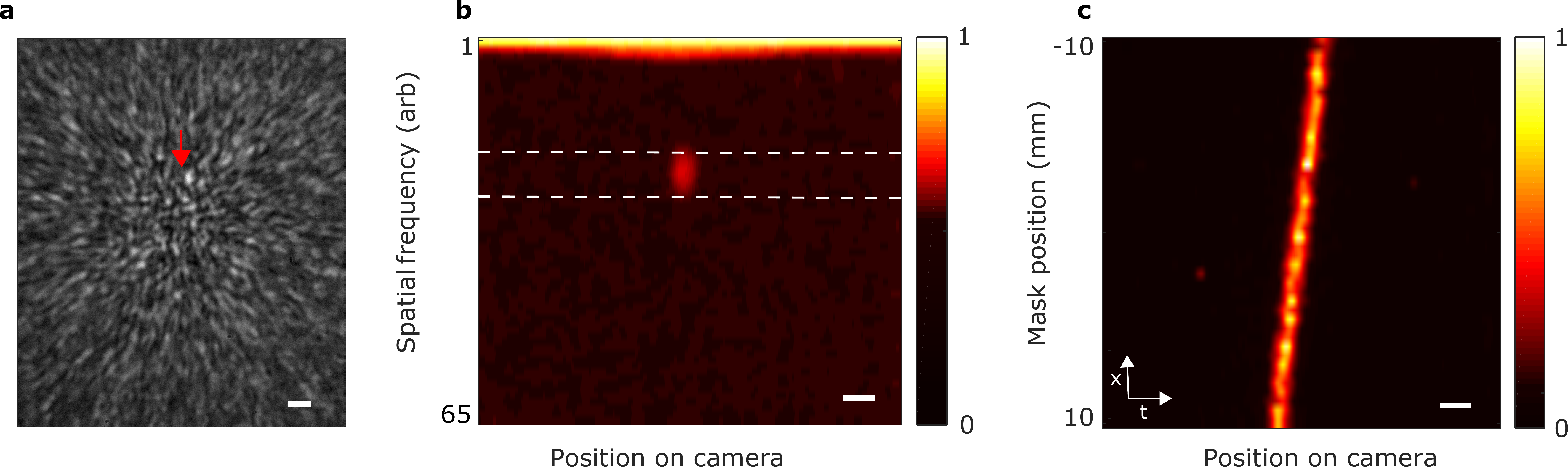}
\caption{\textbf{Extracting the zero delay position using  spectrogram analysis.}\textbf{(a)}, Raw camera image(after median filter).\textbf{(b)}, Spectrogram of the camera image, the blob coincides to the zero delay position,which is the interference peak position.\textbf{(c)},Trace of the interference peak position as a function of the mask position .The trace is obtained from repeating the process shown in \textbf{(A-B)} for different mask positions. Scale bars, 92fs.} 
\label{fig:sup3}
\end{figure}

\section{Localizing sources from hyperbolas intersection}
Several approaches can be used in order to reconstruct the positions of the sources from the spatio-temporal TOF traces (e.g. Supplementary fig. S3c, and Supplementary fig. S4). Such approaches include filtered back-projection \cite{velten2012recovering}, or similar inversion procedures, tracing the light back to ellipsoids \cite{gariepy2016detection} or spheres \cite{o2018confocal}. In our approach, each fringes intensity peak in the spatio-temporal (x-t) trace (e.g. Supplementary fig. S3c) provides a TOF \textit{difference}, and thus localizes the sources on a hyperbola.

To demonstrate a simple localization approach, we have implemented a back-projection procedure, which sums up the contributions of all the hyperboloids retrieved from each peak of the fringes detected intensity. 
To account for the experimental setup mounting inaccuracies, the fringe position to TOF delay was determined based on a set of calibration measurements with set of sources at known positions. Example results of such calibration measurements over a field of view of 70cm x 20mm (z,x) are shown in Supplementary fig. S4. For a coarse calibration, measurements for a set of 63 source positions (7 x 9, in (x,z) respectively) were taken, when each curve contains 40 different mask positions. 
For noise filtering and smoothing the results the reconstructed hyperbolas were convoluted with a rectangular smoothing kernel.
For the localization around a corner (Fig. 4), due to the roughness of the wall surface, the fringes peaks positions varied considerably more as a function of the mask position (Fig.4c), compared to the experiments with the diffusive barrier. In order to reduce the resulting errors in the measured TOF, the hyperboloids were reconstructed from the average fringes positions over 5 adjacent mask positions.

\begin{figure}[H]
\centering
\includegraphics[width=12cm]{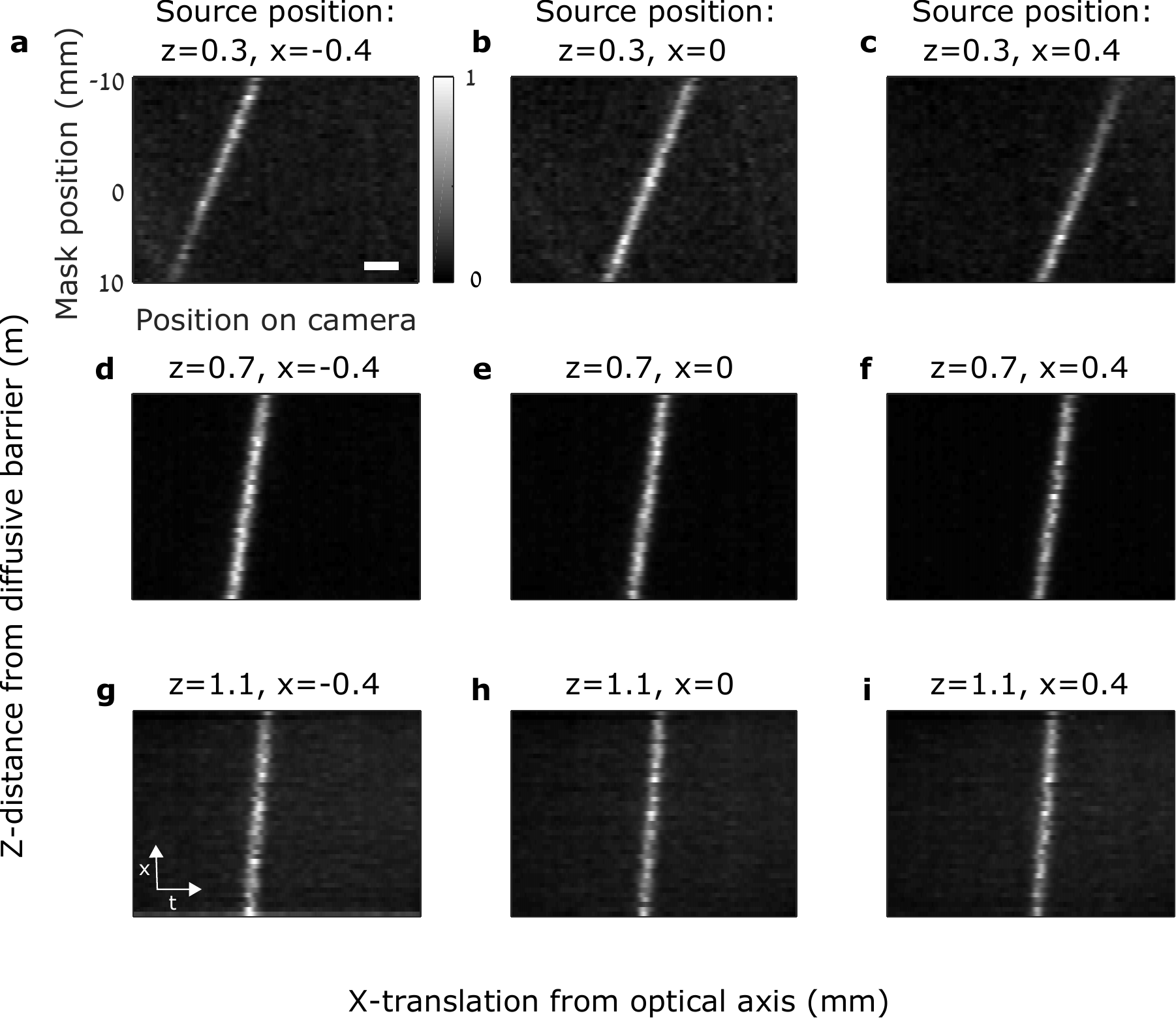}
\caption{\textbf{Spatio-temporal TOF traces for different source positions, used for calibration.} The slope of the trace reflects the source's distance from the diffuser (z position). Sources that are located at a small distance from the barrier (\textbf{A-C}) result in a tilted curve,  while far away sources (\textbf{G-I}) results in a nearly vertical trace.  The traces horizontal (x) position reflects the horizontal (x) position of the source. Scale bars, 92fs. }  
\label{fig:sup4}
\end{figure}

\section{TOF resolution and system geometry consideration}
\subsection{TOF resolution}
In our approach the temporal TOF resolution is given by the envelope of the field cross-correlation (interference of short coherence light). The temporal extent of this envelope is the coherence time of the source, which, according to the Wiener-Khinchin theorem is Fourier-related to the source power spectrum.  The coherence time of the source can thus be estimated from the spectral width.

The power spectrum of the white light LED source used in our experiments is shown in Supplementary fig. S5.   The spectral FWHM width is approximately 
  $\Delta \lambda\approx 140nm$ and the central wavelength is $\lambda_c\approx 600nm$. The coherence time is inversely proportional to the spectral width, $\Delta f$, i.e.:
\begin{equation}
\tau_{c}\approx \frac{1}{\Delta f}  \approx \frac{c \Delta \lambda} {\lambda_c^{2}} \approx 11 fs
\end{equation}
This gives a resolution of the order of 10 fs, three orders of magnitude better than ultrafast detectors used in conventional TOF techniques, which are of the order of 15ps using a streak camera \cite{velten2012recovering}, and 8ps using SPAD detectors. \cite{o2018confocal}
\begin{figure}[H]
\centering
\includegraphics[width=12cm]{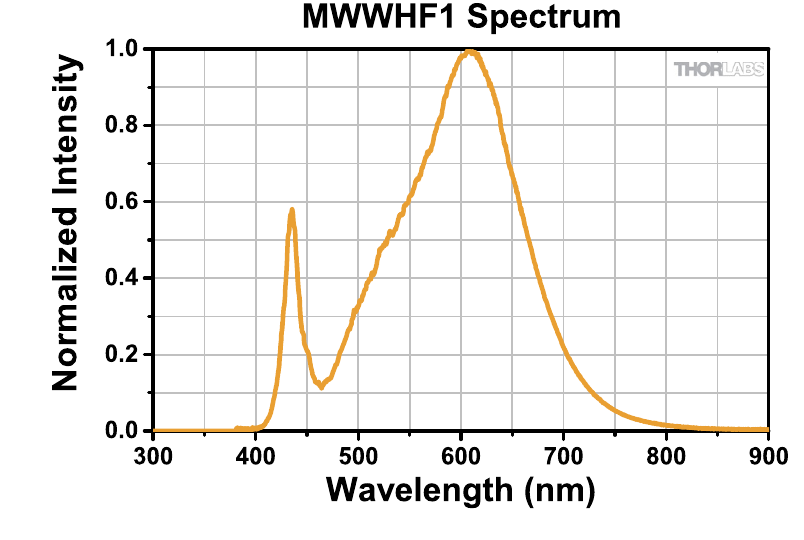}
\caption{\textbf{Spectrum of the white light LED source used in our experiments (MWWHF1 data-sheet, Thorlabs)}.} 
\label{fig:sup5}
\end{figure}
\subsection{Fulfilling the spatial Nyquist sampling criteria}
In order to be able to measure the interference fringes, the interference pattern has to be spatially Nyquist sampled, i.e. the camera pixel pitch $\Delta x_{pixel}$ has to be smaller than half the spatial period of the white light fringes $\Lambda/2$. This period is dictated by the central wavelength of the light source, $\lambda_0$, and the geometry of the measurement system (Supplementary fig. S6):
In our setup, the interference fringes are a result of interference between two apertures separated by a distance $D=3.2mm$. The aperture mask is placed at the front focal plane of a lens ($f=100 mm$), and the camera is positioned in the back focal plane of the lens. In this geometry the spherical wave that is transmitted through each of the slits becomes a plane wave propagating at an angle $\theta=atan(D/2f)$ after passing through the lens. The low coherence interference fringes have a period which is $\Lambda=\lambda_0/2sin(\theta) \approx \frac{\lambda_0  f}{D}$. In our setup $D=3.2 mm$ and $f=100 mm$ were chosen such that $\Lambda \approx 19 \mu m$ , which is approximately three times larger than the camera pixel pitch $\Delta x_{pixel}= 6.5 \mu m$ , fulfilling the required Nyquist sampling criterion.

\begin{figure}[H]
\centering
\includegraphics[width=8cm]{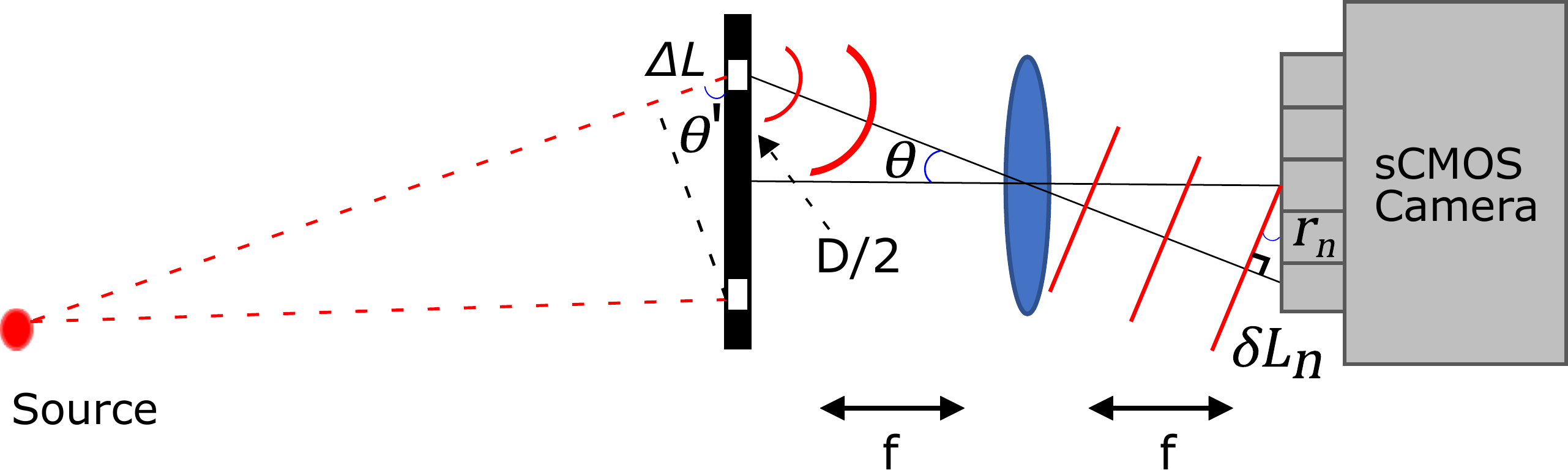}
\caption{\textbf{System geometry and path length difference $\delta L $ to the different camera pixels}} 
\label{fig:sup6}
\end{figure}

\subsection{Camera pixel to TOF difference conversion}
The calculation of the optical path length difference, $\delta L$ (i.e. the time delay times the speed of light, $\delta L = c\Delta t$) for each pixel in the camera plane is depicted in Supplementary fig. S6.  The position of the fringes on the camera, $r_n$, is directly converted to the TOF delay by considering that the relative path length difference from the on-axis camera pixel to the $n^{th}$ pixel is $\delta L_n = r_n sin(\theta)$ , where $r_n=\Delta x_{pixel}\cdot n$ is the position of the $n^{th}$ pixel. 
Thus, the time delay corresponding to fringes measured at  the $n^{th}$ pixel position is given by:
\begin{equation}
\Delta T_n=2\delta L_n / c = 2r_n\cdot \sin(\theta)/c\approx\frac{r_n\cdot D}{c\cdot f}
\end{equation}
In the last step we assumed $f\gg D,r_n$, as is the case in our experiments. 

Substitution of our setup chosen parameters: $f=100mm$, $D=3.2mm$, and $r_n =n\cdot 6.5\mu m$, will give the time delay for the $n^{th}$ pixel: 
\begin{equation}
\Delta T_n\approx n\cdot 0.6fs
\end{equation}
resulting in a conversion of each camera pixel in our spatio-temporal maps to a TOF delay of $0.6fs$, roughly a third of the central wavelength, as required for proper Nyquist sampling.

\section{Spatial localization resolution for a finite sized object}
Here we provide an estimate of the \textit{spatial} localization resolution provided by the TOF \textit{temporal} measurement resolution. 
For a distant object, located at an angle $\theta$ with respect to the barrier (Supplementary fig. S6), the TOF difference of the light to two apertures at a separation $D$  is
\begin{equation}
\Delta t \approx D\cdot \cos (\theta ')/c
\end{equation}
This TOF difference is measured with a resolution of the source coherence time: $\tau_c =d(\Delta t)\approx l_c/c$, where $l_c$ is the source coherence length. 
The dependence of the spatial localization on the temporal measurement can be estimated by taking the derivative of $\Delta t$   with respect to theta, yielding: 
\begin{equation}
d(\Delta t)=D\cdot\sin\theta\cdot d\theta/c
\end{equation}
Thus, plugging $d(\Delta t)\approx l_c/c$, the angular localization resolution from a single camera shot is approximately :
\begin{equation}
d\theta'\approx l_c/(D\cdot \sin(\theta')) \approx 1.5 mrad
\end{equation}
Where in the last step we substituted the parameters of our experiment $l_c\approx 3.3\mu m$ , and $D=3.2mm$, at $\theta'=45\deg$.
The spatial resolution is improved for larger apertures separation, $D$ (i.e. a larger base-length). However, for an extended source, in order to obtain high contrast fringes, the apertures separation must be smaller than the coherence size of the source at the barrier, $r_{coh}$.  According to the Van-Cittert Zernike theorem,  $r_{coh}\approx z\lambda /D_{obj}$, where $z$ is the object distance from the barrier and $D_{obj}$ is the object's diameter. The largest apertures separation providing high contrast fringes is thus of the order of the coherence size: $D\approx r_{coh}$. This will yield an angular resolution from a single camera shot of:
\begin{equation}
d\theta'\approx l_c/(r_{coh}\cdot \sin(\theta')) \approx l_c\Delta \theta_{obj}'/(\lambda\cdot \sin(\theta))
\end{equation}
where $\Delta \theta_{obj}' = D_{obj}/z$ is the angular size of the hidden source. 

For broadband white-light sources the coherence length, $l_c$ is of the order of the wavelength, $\lambda$ . Thus, the single-shot localization resolution is fundamentally limited to be comparable to the object angular size. 

The final localization accuracy is considerably better than the single shot localization resolution, since it is obtained from \textit{multiple} single-shot TOF measurements at different masks positions, effectively yielding a large base-length.

\section{Characterization of white-painted wall reflection}
\begin{figure}[H]
\centering
\includegraphics[width=12cm]{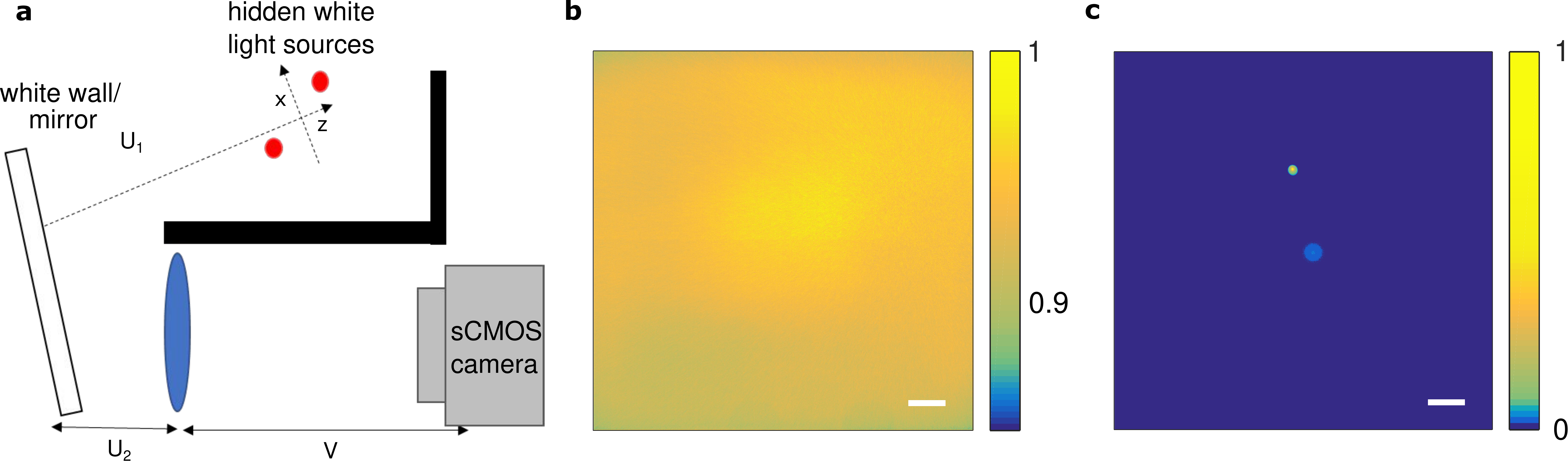}
\caption{\textbf{Impossibility for conventional imaging around the corner:}\textbf{(a)} Conventional imaging setup: the two light sources of Fig.4 are placed around the corner. A single lens images the plane between the two points (located at a distance $U_1 + U_2$ from the lens) on the camera, located at a distance $V$ from the lens. \textbf{(b)} The image obtained from light reflected by the white diffusive wall is featureless, showing no specular reflection. \textbf{(c)} The image obtained with a mirror replacing the wall, showing the two light sources. Scale bars, 1cm. }
\label{fig:sup7}
\end{figure}

In the  around the corner localization experiments of Fig. 4, the light was collected at a reflection angle approximately equal to the angle of incidence of the light from the object. This was done in order to maximize the light collection efficiency, and speckle contrast. The latter is maximized as the speckle spectral decorrelation width is maximized, which occurs in this angle \cite{small2012spectral}. In order to verify that there is no significant specular reflection at this angle that can reveal the position of the light sources with conventional imaging, we have performed conventional imaging with the light reflected from the wall, and compared to the case when the wall was replaced by a mirror. 

The results of these measurements are displayed in Supplementary fig. S7.  The setup (Supplementary fig. S7a) is a simple single lens imaging setup, which images the two hidden objects of Fig. 4a, on a camera.  The light source that is closer to the wall is located at a distance $U_1 + U_2=56+14=70cm$ from the lens ($f=8cm$), and an sCMOS camera is positioned at a distance $V=9cm$ behind the lens, such that $(U_1+U_2)^{-1} + (V)^{-1} = f^{-1}$ .

Supplementary fig. S7b shows the image recorded using the light reflected from the white painted wall, showing no information on the hidden objects positions. In contrast, replacing  the wall with a mirror wall clearly reveals the positions of the two light sources  (Supplementary fig. S7c).

\section{Influence of a thick barrier on the temporal cross-correlation}
In this section we theoretically analyze the potential distortions induced by multiple-scattering in a thick barrier on the measured temporal cross-correlation. Consider two fields $E_1(t)$ and $E_2(t)$, arriving at the diffusive barrier at two different points, as depicted in Supplementary fig. S8. Each of the fields exiting the barrier $E_{i,m}$ , which are measured by our system, are given by the convolution of the entering field with the impulse response function of the barrier, $h_i(t)$, for the specific input-output positions $r_i$, such that:
\begin{equation}
E_{1,m}\left(t\right)	=E_{1}\left(t\right)\ast h_{1}\left(t\right)
\end{equation}
\begin{equation}
E_{2,m}\left(t\right)	=E_{2}\left(t\right)\ast h_{2}\left(t\right)
\end{equation}\begin{figure}[H]
\centering
\includegraphics[width=6cm]{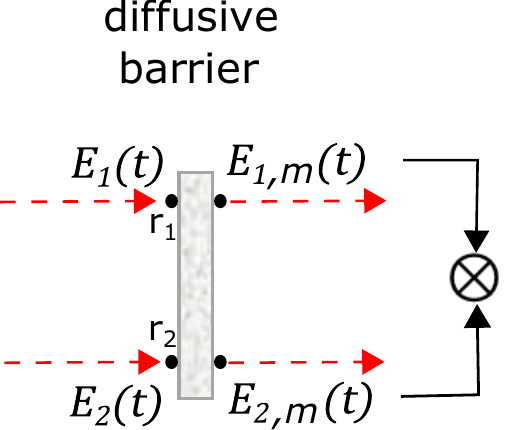}
\caption{\textbf{The influence of the impulse response of the barrier} }
\label{fig:sup8}
\end{figure}

The cross-correlation of the measured fields exiting the barrier is thus given by:
\begin{equation}
E_{1,m}\otimes E_{2,m}\left(t\right)=[E_{1}\left(t\right)\ast h_{1}\left(t\right)] \otimes [ E_{2}\left(t\right)\ast h_{2}\left(t\right)]
\end{equation}
Since all the fields and impulse responses are real functions, the cross-correlation of the convolutions is equal to the convolution of the autocorrelations:
\begin{equation}
E_{1,m}\otimes E_{2}\left(t\right)=[E_{1}\left(t\right)\otimes E_{2}\left(t\right)]\ast [h_{1}\left(t\right)\otimes h_{2}\left(t\right)]
\end{equation}
 For a sufficiently thin barrier 
the impulse responses, $h_i(t)$, can be approximated as delta functions, thus providing a good estimate for the arriving fields cross-correlations, as required:
\begin{equation}
E_{1,m}\otimes E_{2,m}\left(t\right)\approx E_{1}\left(t\right)\otimes E_{2}\left(t\right)
\end{equation}
In the case of thick multiply scattering barriers, the measured cross-correlation will be the convolution of the desired cross-correlation with the cross-correlation of the impulse responses, which is a function that is limited in time to twice of the Thouless time (dwell time) of the light in the medium. Thus, a thick scattering barrier will induce distortions and smearing that are given by the path delay spread of the light in the medium. While effectively lowering the temporal resolution, the temporal cross-correlation approach is effective also through effectively thick barriers, as we prove experimentally in the results presented in Figure 4. 

\end{document}